\documentclass[twocolumn,showpacs,preprintnumbers,amsmath,amssymb]{revtex4}
\usepackage{graphicx}
\usepackage{dcolumn}
\usepackage{bm}

\newcommand{\bef}{\begin{figure}}
\newcommand{\eef}{\end{figure}}
\newcommand{\nn}{\nonumber}
\newcommand{\be}{\begin{equation}}
\newcommand{\ee}{\end{equation}}
\newcommand{\bea}{\begin{eqnarray}}
\newcommand{\eea}{\end{eqnarray}}
\begin{document}

\title{Lepton interferometry in relativistic heavy ion collisions - a 
case study}

\author{Jan-e Alam$^1$, Bedangadas Mohanty$^1$, A. Rahaman$^2$, Sourav 
Sarkar$^{1,}$\footnote{ Presently at: IFIC, University of Valencia, Aptdo. 
Correos 22085, 46071 Valencia, Spain} and Bikash Sinha$^{1,3}$}

\medskip

\affiliation{$^1$Variable Energy Cyclotron Centre, 1/AF Bidhan Nagar, 
Kolkata 700064, India\\
$^2$Durgapur Government College, Durgapur, West Bengal, India,
$^3$Saha Institute of Nuclear Physics, 1/AF Bidhan Nagar, Kolkata 700064, India}

\date{\today}

\begin{abstract}
We propose intensity interferometry with identical lepton pairs
as an efficient tool for the estimation of the source size of the
expanding hot zone produced in relativistic heavy ion collisions.
This can act as a complementary method to two photon interferometry.
The correlation function of two electrons with the same helicity 
has been evaluated for RHIC energies. The thermal shift of the $\rho$
meson mass has negligible effects on the HBT radii.
\end{abstract}

\maketitle

\section{INTRODUCTION}

  In high energy heavy ion collisions two-particle 
  correlations have been extensively studied both 
  experimentally~\cite{na35,hb7,hb8} and theoretically~\cite{hb3,uaw,tcsorgo}
  to obtain direct information about the size, 
  shape and dynamics of the source at freeze-out. This is 
  usually done via selection of transverse momentum and rapidity of 
  the correlated particles. Such calculations are based on the fact that
  identical particles occurring nearby in phase-space experience
  quantum statistical effects resulting from the (anti)symmetrization
  of the multi-particle fermion (boson) wave function. 
  For bosons, the two-particle
  coincidence rate shows an enhancement at small momentum difference
  between the particles and an opposite behaviour is observed for
  fermions. The momentum range of this enhancement or depletion 
  can be related to the size of the particle source in co-ordinate
  space.

  Recently, it has been argued that 
  Hanbury-Brown and Twiss (HBT) interferometry is a sensitive probe
  of the QCD equation of state and hence formation of QGP~\cite{hb6,hb9,hb10}.
  Further, it has been argued that in contrast 
  to hadrons, two-particle intensity 
  interferometry of photons~\cite{photon,dks,DP,WA98} 
  which are produced throughout the space-time 
  evolution of the reaction and which suffer almost no interactions with 
  the surrounding medium can provide information on the 
  the history of the evolution of the hot 
  matter created in heavy ion collisions~\cite{JA}. 

The two fermion interferometry is a well known  method
used in nuclear physics to estimate the size of the
source formed after nuclear collisions~\cite{AL}(see~\cite{alexander} for
a review). In case of fermion interferometry
the symmetric (anti-symmetric) space part of the wave function is 
coupled with the anti-symmetric (symmetric) spin part of the wave
function. In this work  we will evaluate correlation function
for the symmetric spin part and hence anti-symmetric space wave functions.
Statistical spin mixture has been neglected here~\cite{alexander}.

In our previous work on photon interferometry~\cite{photon} the spin-averaged
source function was used to study the two photon correlation functions.
Here we use spin-dependent source functions to evaluate the 
two-lepton correlation function.  The 
main aim of this study is to extract the HBT radii of the
source from spin-dependent electron correlation function
 and
show that the HBT radii extracted
here is similar to those obtained from the two
photon interferometry. This indicates that the results from
photon interferometry is not very sensitive to the 
spin of the photon. 
  In addition to this it will be useful to see the 
effects of the shift of vector meson masses with temperature
on the correlation function of two electrons.

 The paper is organized as follows. In the next section we give a
 general discussion on the correlation function for fermions and 
 the associated kinematics. This is followed by the section which 
 deals with the space time evolution. 
 Section~IV, deals with the results on two lepton correlations 
 at RHIC energies, results with  mass variation of vector mesons in a 
 hot medium and brief discussion on the results of two electron 
 interferometry vis-a-vis two photon interferometry. We summarize our 
 findings in the last section.

\section{CORRELATION FUNCTION}

The two-particle correlation function in momentum space
is defined as,
\begin{equation}
C_{2}(\vec{k_{1}}, \vec{k_{2}}) = \frac{P_{2}(\vec{k_{1}}, \vec{k_{2}})}
{P_{1}(\vec{k_{1}}) P_{1}(\vec{k_{2}})} 
\label{eq1}
\end{equation}
where $\vec{k_1}$ and $\vec{k_2}$ are the three momenta of the two particles.
$P_{1}(\vec{k_{i}})$
and $P_{2}(\vec{k_{1}}, \vec{k_{2}})$ represent 
the one- and two- particle inclusive
electron spectra respectively. These are defined as,

\begin{equation}
P_{1}(\vec{k}) = \int d^{4}x~S(x,k)
\label{eq2}
\end{equation}
and
\begin{widetext}
\begin{eqnarray}
P_{2}(\vec{k_{1}}, \vec{k_{2}}) = P_{1}(\vec{k_{1}}) P_{1}(\vec{k_{2}})  -
\int d^{4}x_{1} d^{4}x_{2} ~S(x_{1},K) 
~S(x_{2},K)~\cos(\Delta x^{\mu} \Delta k_{\mu}) 
\label{eq3}
\end{eqnarray}
\end{widetext}
where, $K=(k_1+k_2)/2$, $\Delta k_\mu=k_{1\mu}-k_{2\mu}=q_\mu$,
$x_i$ and $k_i$ are the four-coordinates of position and momentum 
respectively and $S(x,k)$ is the source function, which  defines
the average number of electrons with four-momentum $k$ emitted from a 
source element centered at the space-time point $x$. In the
present case $S(x,k)$
is the thermal emission rate of electrons 
per unit four volume. For processes of the form 
$\alpha(p_1) + \beta (p_2)\,\rightarrow
e^+(k_1)\,+\,e^-(k)$, we have~\cite{annals},
\begin{eqnarray}
S(x,k)&=&\frac{{\cal{N}}}{16(2\pi)^7k}\,\int_{(m_1+m_2)^2}^{\infty}
\,ds\,\int_{t_{{\rm min}}}^{t_{{\rm max}}}\,dt\,|{\cal M}|^2\nonumber\\
&&\times\int\,dE_1
\int\,dE_2\frac{f(E_1)\,f(E_2)}{\sqrt{aE_2^2+
2bE_2+c}},
\label{sorce}
\end{eqnarray}
where $\alpha$ and $\beta$ are either quarks or pions, 
$k_i=(E_i,\vec{k_i})$ is the four vector for the
particle $i$. The masses of quarks and electrons are
neglected here.
${\cal N}$ is the overall degeneracy of the particles $\alpha$ and $\beta$,
$f(E_i$)  denotes
the thermal distribution functions and $s$, $t$, $u$ are the 
usual Mandelstam variables. The expressions for $a$, $b$, $c$,
and the integration limits, $E_{1min}$, $E_{2min}$ and $E_{2max}$ 
are given in Ref.~\cite{annals}.

The spin-dependent invariant amplitude, ${\cal M}$,
for the processes
$q_R^-\,q_L^+\rightarrow e_R^- e_L^+\,( e_L^- e_R^+)$,
$q_L^-\,q_R^+\rightarrow e_R^- e_L^+\,( e_L^- e_R^+)$   
and $\pi^-\pi^+\rightarrow e_R^- e_L^+\,( e_L^- e_R^+)$  have been
calculated using standard field theoretic techniques~\cite{peskin}.

We shall be presenting the results as a function of
longitudinal ($q_{long}$),
outward ($q_{out}$), sideward ($q_{side}$) 
and invariant momentum differences ($q_{inv}$) of the two leptons 
and these are defined as,
\bea
q_{long} & = &k_{1z}-k_{2z}\nn\\
	 & = &k_{1T}\sinh(y_1)-k_{2T}\sinh(y_2)
\label{eq8}
\eea
\bea
q_{out}& = & \frac{\vec{q}_{T} \cdot \vec{K}_{T}}{\left|{K_{T}}\right|}\nn\\
       & = & \frac{(k_{1T}^{2} - k_{2T}^2)}{
\sqrt{k_{1T}^{2} + k_{2T}^{2} + 2 k_{1T}k_{2T} \cos(\psi_{1}-\psi_{2})}}
\label{eq5}
\eea
\bea
q_{side}& = &{\left| \vec{q}_{T} - q_{out} \frac{ \vec{K}_{T}}{K_{T}}\right|}
\nn\\ 
	& = &\frac{2 k_{1T}k_{2T} \sqrt{1-\cos^{2}(\psi_{1}-\psi_{2})}}{
\sqrt{k_{1T}^{2} + k_{2T}^{2} + 2 k_{1T}k_{2T} \cos(\psi_{1}-\psi_{2})}}
\label{eq6}
\eea
\bea
q_{inv}^2 & = &{-2 k_{1T}k_{2T} \Bigl[\cosh(y_{1}-y_{2}) - \cos(\psi_{1}
-\psi_{2}) \Bigr]},\nn\\
\label{eq7}
\eea
where, $\vec{q}_{T} = \vec{k}_{1T} - \vec{k}_{2T}$, $\vec{K}_{T} =
(\vec{k}_{1T} + \vec{k}_{2T})/2$ with the subscript $T$ indicating
the transverse component,
$y_i$ is the rapidity and $\psi_i$'s are the angles made 
by $k_{iT}$ with the $x$-axis. 

The HBT radii can be extracted by parametrizing the calculated 
correlation function in the following Gaussian form:
\be
C^{Asym}_2=
1-\lambda\exp(-R^2_{out}q^2_{out}-R^2_{side}q^2_{side}-R^2_{long}q^2_{long})
\label{param1}
\ee
The superscript $`Asym'$ indicates that antisymmetric space wave functions have 
been used. Also note that a negative sign appears before the exponential (as
opposed to a positive sign in case of the two-boson correlation function).
The $\lambda$ in Eq.\ref{param1} is called chaoticity parameter, 
$\lambda=0$ for a coherence source and $\lambda=1$ for a completely chaotic
source. We have set $\lambda=1$ as we are concerned  here with a completely
thermalized system from where the leptons are emitted. If the source
function has more than one maxima then Gaussian 
parametrization fails to describe the structure of the correlation
function at large values of outward, sideward or invariant momenta
defined in section II. For ``binary'' source functions like mixed
phase in the present context, more than one maxima in correlation
functions are possible~\cite{tcsorgo}.
The values of $C^{Asym}_2$ will lie between 0 and 1 for a chaotic source.

A gross idea of the source size can also be obtained from $R_{inv}$ 
which is defined as:
\be
C^{Asym}_2=1-\exp(-R^2_{inv}q^2_{inv})
\label{param2}
\ee

\section{SPACE-TIME EVOLUTION}

For the evaluation of two-electron correlations at RHIC energies
we will consider a  scenario 
where QGP is formed in the initial state 
which evolves with time into a hadronic phase via an 
intermediate mixed phase in a first order phase transition scenario. 
The mixed phase is a mixture of both quark matter and hadronic matter, 
with the fraction of quark matter decreasing with time to zero when the 
phase transition is complete. The hot hadronic gas then expands till the 
system freezes-out. 
We also evaluate the correlation function when
the vector meson masses ($\rho$ in this case) varies with temperature according
to the universal scaling scenario  
proposed by Brown and Rho (BR)~\cite{geb}. 

The initial condition for RHIC energies in terms of the initial 
temperature ($T_{i}$) is calculated from the number of particles per unit 
rapidity at the mid-rapidity region for those 
energies according to the following equation,
\be
T_i^3=\frac{2\pi^4}{45\zeta(3)}\frac{1}{\pi R_A^2\tau_i 4a_k}\,
\frac{dN}{dy}
=\frac{1}{\pi R_A^2\tau_i 4a_k}\,
\frac{dS}{dy}
\ee
where $dS (dN)$ is the entropy (number) contained within a volume 
element $\Delta\,V=\pi\,R^2\,\tau_i\,dy$, $R$ as the radius of 
the colliding nuclei. $g_{eff}$ is the effective statistical degeneracy,
$\zeta(3)$ denotes the Riemann zeta function
and $y$ is the rapidity. For massless  bosons
(fermions) the ratio of $dS$ to $dN$ is given by 
${2 \pi^4}/({45 \zeta(3)})\,\sim\,3.6$ (4.2), which is a crude
approximation for heavy particles. 
For example the above ratio is 3.6 (7.5) for 140 MeV pions (938 MeV protons)
at a temperature of 200 MeV.
$a_k=\pi^2g_k/90$ is determined by the statistical degeneracy ($g_k$)
of the system formed after the collision.
Taking the particle multiplicity per unit rapidity 
to be 1100 for Au+Au collisions at RHIC, 
we get $T_{i}$ = 264 MeV for an initial time  
of 0.6 fm/c~\cite{dumitru}. The critical temperature ($T_{c}$) 
is taken to be 170 MeV~\cite{karsch} here.
The freeze-out temperature ($T_{f}$) 
is taken to be 120 MeV, which describes the transverse momentum 
distributions of produced hadrons~\cite{bkp} with the same
space time evolution model used here. 
For the equation of state (EOS) which plays a central role in
the space time evolution, we have considered a 
hadronic gas with particles of mass up to
2.5 GeV and including the effects of non-zero widths of various mesonic 
and hadronic degrees of freedom~\cite{bm2}. 
The velocity of sound ($c_s$) corresponding to this EOS at freeze-out 
is about 0.18~\cite{bm2}. The bag model EOS has been used for the QGP phase.
Using  the above inputs  and assuming 
 boost invariance along the longitudinal direction~\cite{bj}
the (3+1)-dimensional hydrodynamic equations 
have been solved to study the space time evolution~\cite{hvg}
from the initial QGP phase to freeze-out with
an intermediate mixed phase of QGP and hadrons.  The initial
energy density profile used here is same as the one used in~\cite{prcr}
and the initial radial velocity is taken as zero.

\section{RESULTS}

In this section we present the results of two-electron interferometry
at RHIC energies. In Fig.~\ref{fig0} the transverse momentum distribution
of the electrons for Au + Au collisions are depicted for
various phases. The initial temperature is taken as $T_i=264$ MeV. 
We have assumed that 
the mass of the intermediary $\rho$ in pion annihilation
varies according to Brwon-Rho scaling~\cite{geb}.

\bef
\begin{center}
\includegraphics[scale=0.4]{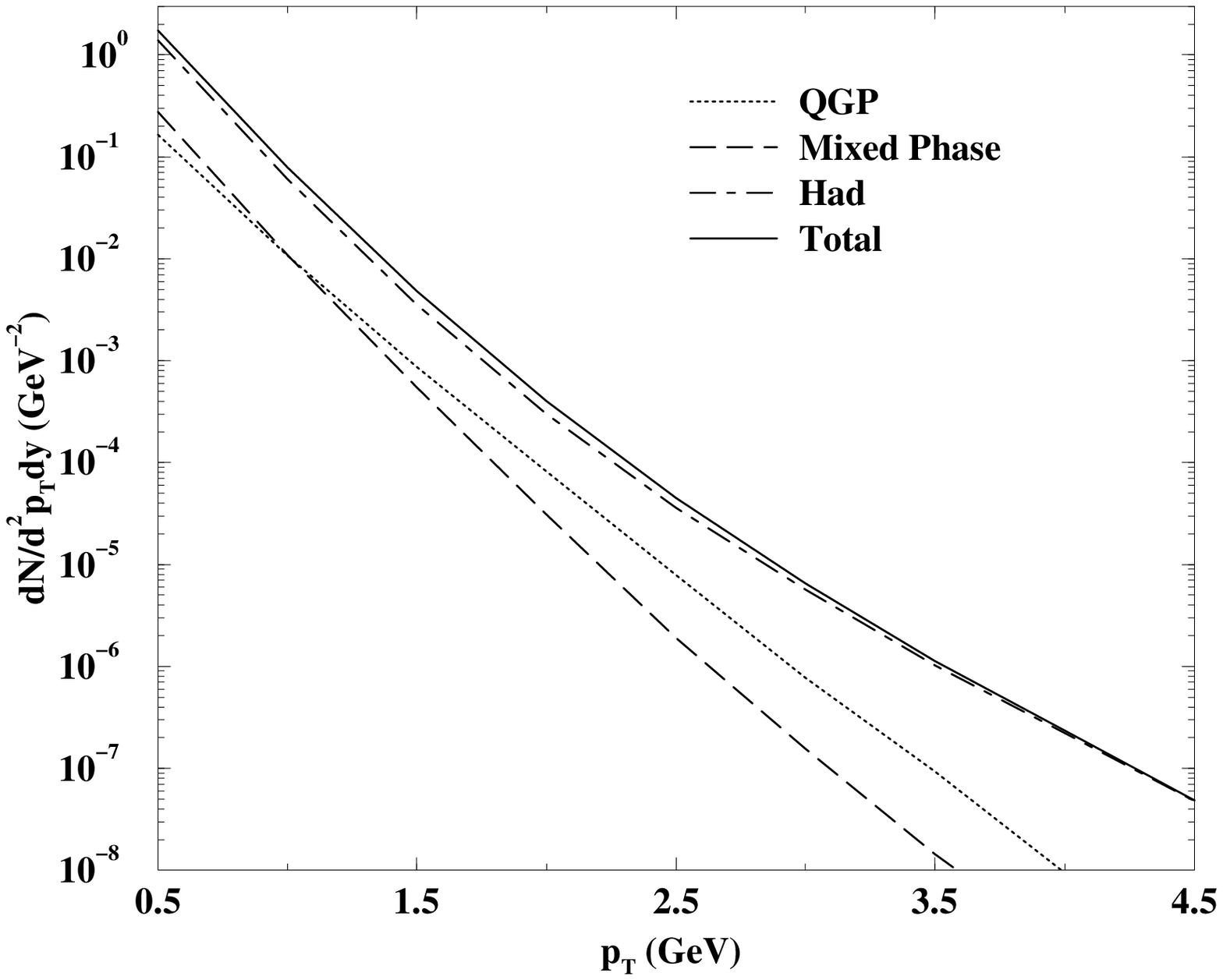}
\caption{Transverse momentum distribution of single electron 
for Au + Au  collisions at 200 AGeV at RHIC. Dotted (dashed)
line indicate results for QGP (Mixed) phase and dot-dashed (solid) line 
represent $p_T$ spectra for hadronic (total) phase.
}
\label{fig0}
\end{center}
\eef


The results on the interferometry can be presented for several 
combinations in the variables $y$, $\psi$ and $k_{T}$.  
For simplicity, we will present all 
two-lepton correlation functions for single electrons with momentum 
1 GeV/c, with the assumption that around this value of transverse momentum
most of the electrons will have a thermal origin.
We take $\psi_{2} = 0$ and $y_{2} = 0$ for all cases 
varying $\psi_{1}$ and $y_{1}$ wherever necessary.
\bef
\begin{center}
\includegraphics[scale=0.4]{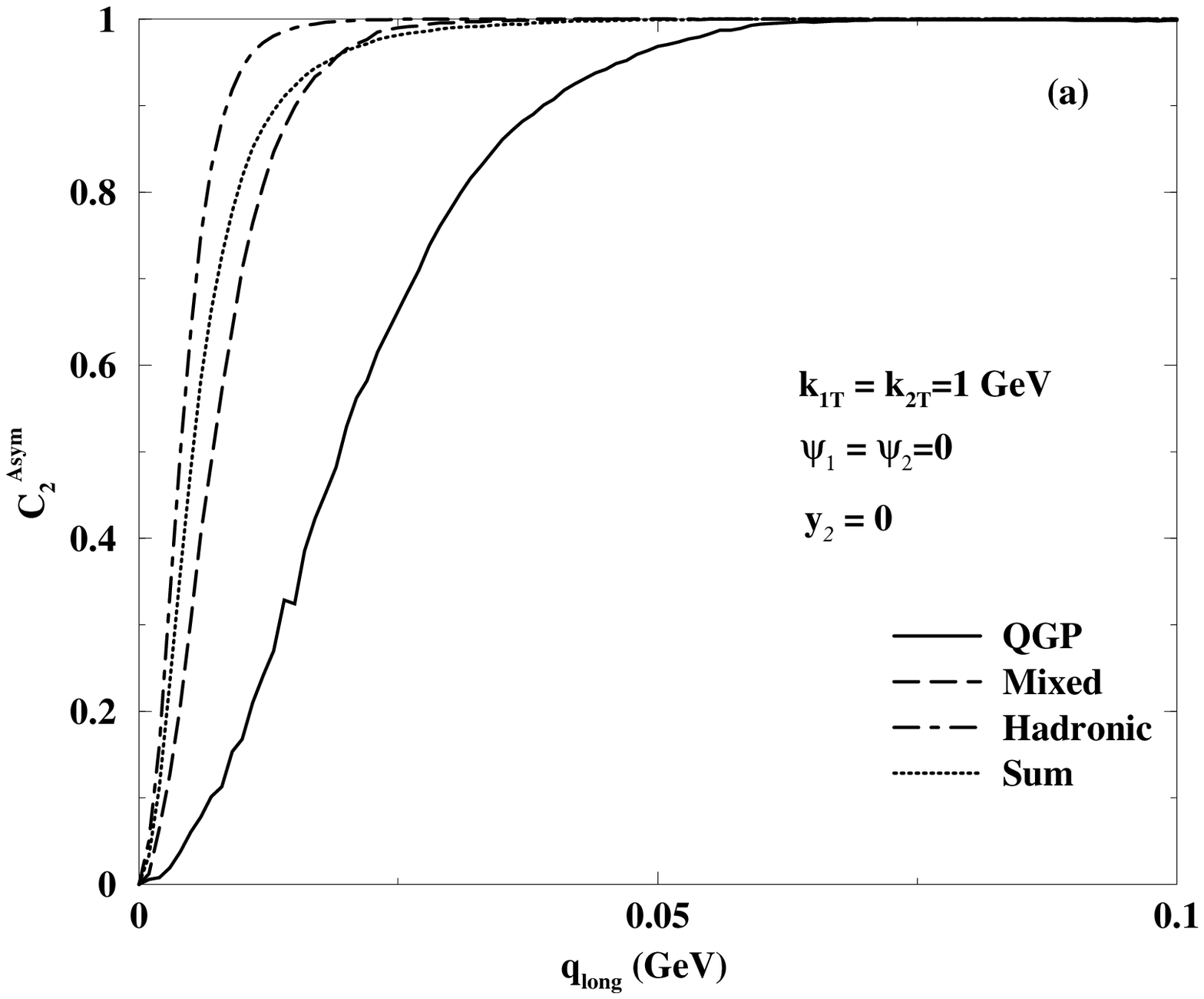}
\includegraphics[scale=0.4]{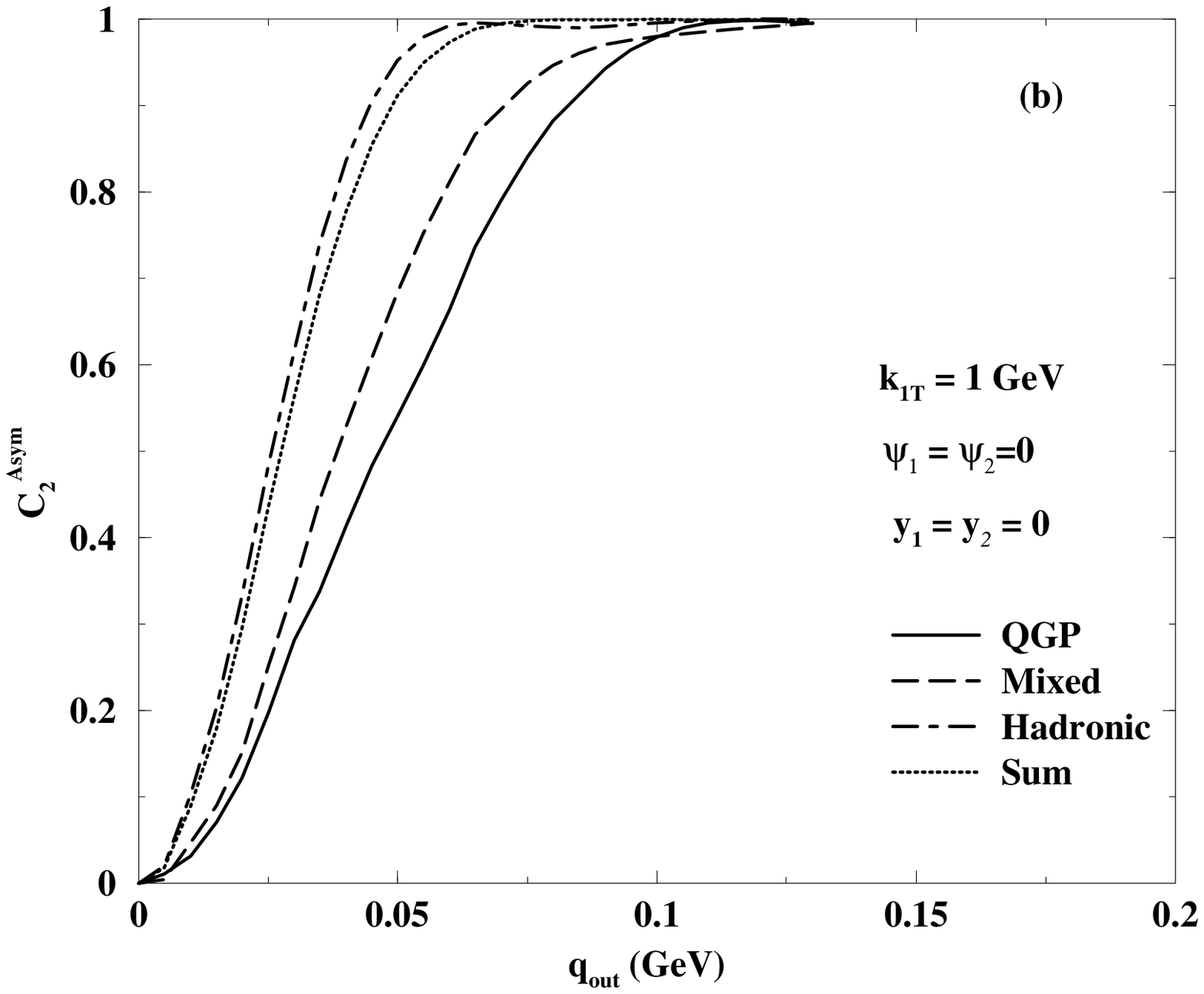}
\includegraphics[scale=0.4]{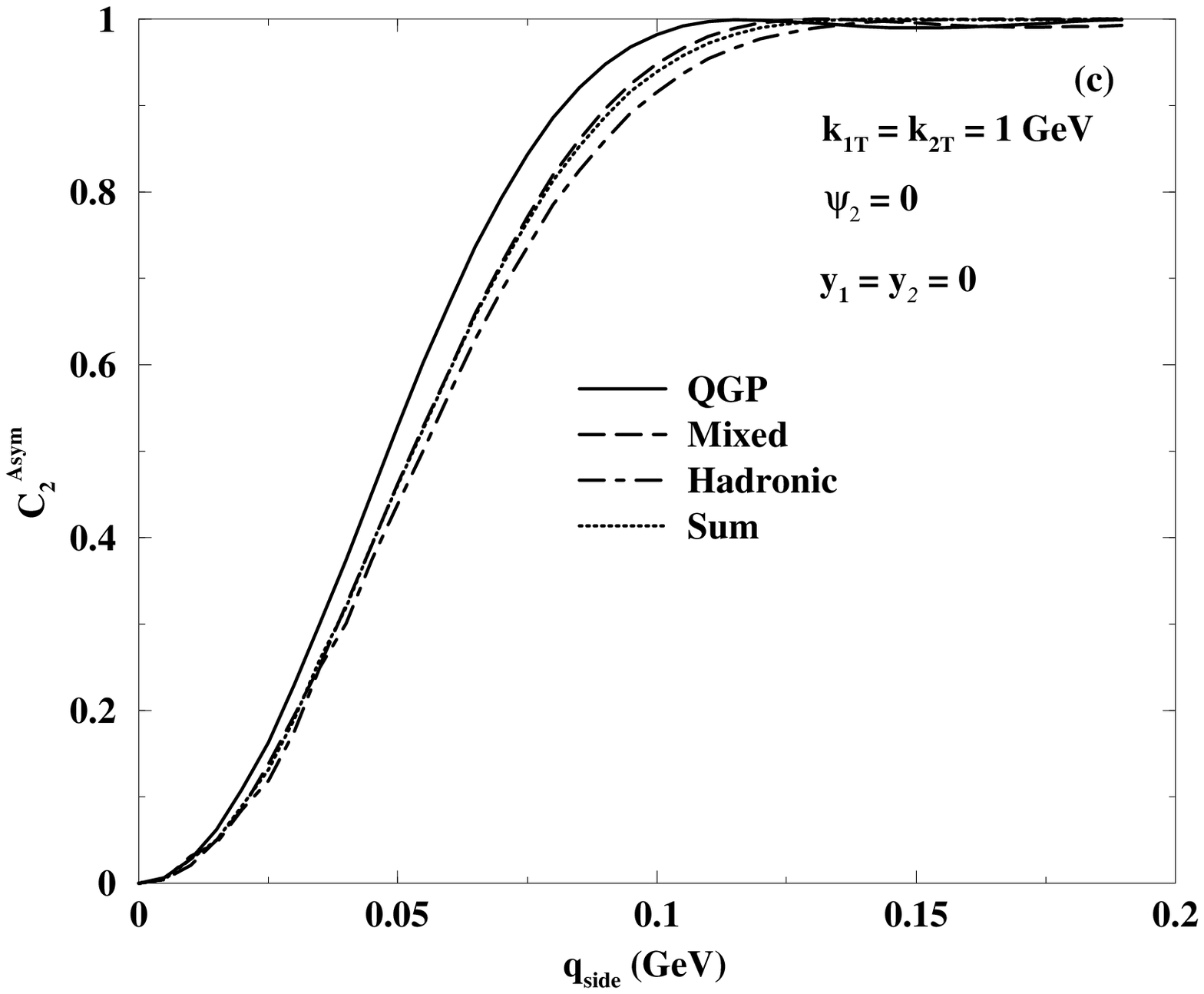}
\includegraphics[scale=0.4]{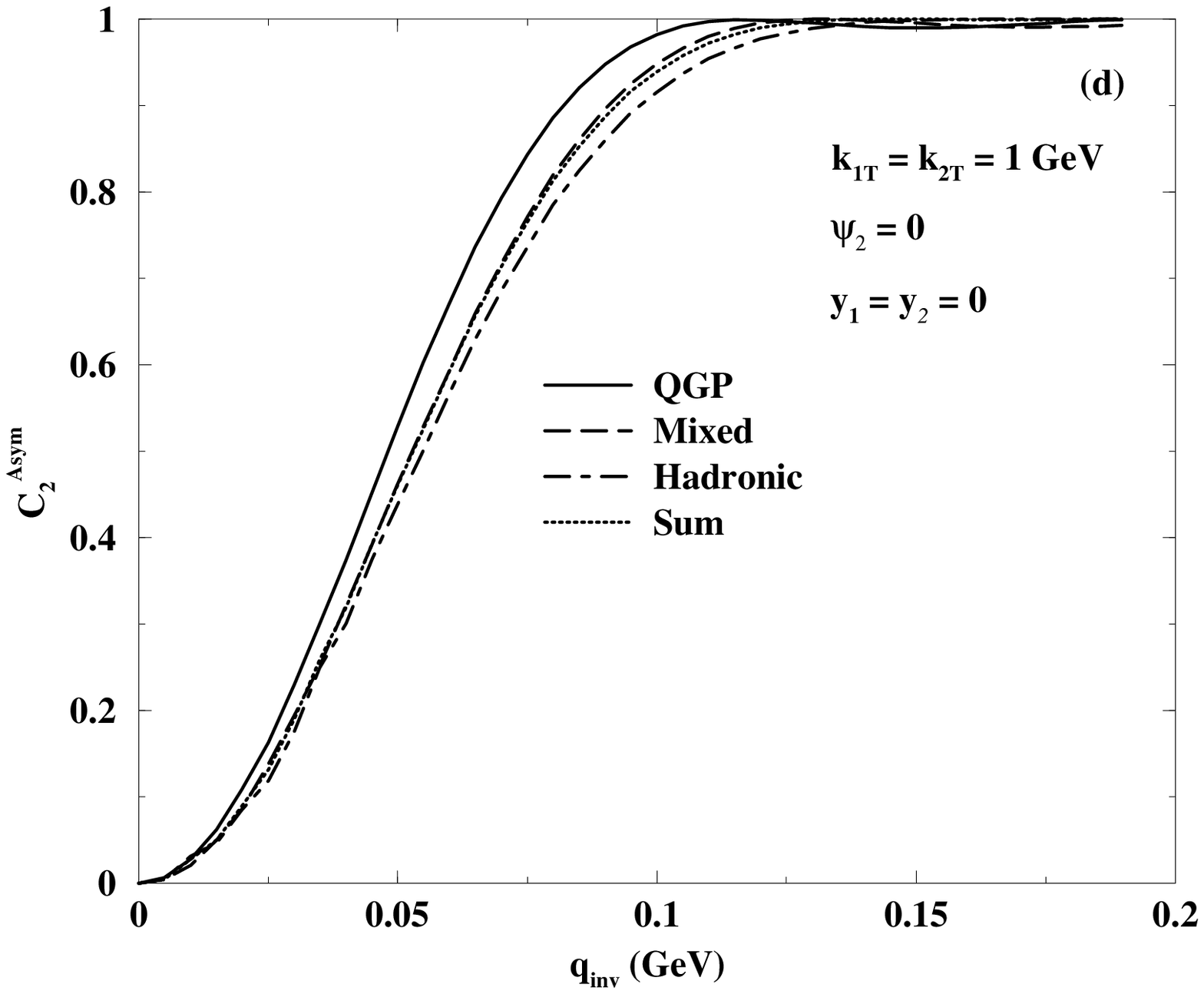}
\caption{Correlation function, $C^{Asym}_2$ as a function of $q_{long}$,
$q_{out}$, $q_{side}$ and $q_{inv}$
for Au + Au  collisions at 200 AGeV at RHIC. Solid (dashed)
line indicate results for QGP (Mixed) phase and dot-dashed (dotted) line 
represent correlation function for hadronic (sum) phase.
}
\label{fig1}
\end{center}
\eef
Figs.~\ref{fig1}(a-d) show the variation of the correlation 
strength ($C^{Asym}_{2}$) as a function of $q_{long}$, $q_{out}$, $q_{side}$
and $q_{inv}$ for various phases (sum $\equiv$ QGP+Mixed+hadrons). 
The HBT dimensions extracted from these correlation functions
are shown in Table I. To obtain the longitudinal dimension $R_{long}$
in the longitudinally co-moving system (LCMS) of reference one
should multiply the numbers given in Table I by the Lorentz factor
$\gamma_K\,(=coshy_K)$, where
$y_K$ is the rapidity corresponding to $K$ defined in section II.
The width of the correlation function for the
hadronic phase is the largest as compared to the other
phases along $q_{out}$, followed by that for the mixed phase and the
QGP phase. For $q_{side}$, the values are comparable for QGP and
mixed phases, while it is slightly lower for hadronic phase.
The HBT dimensions satisfy the relation 
$R_{out}/R_{side}\,\sim\,2$ for
the correlation functions denoted by `sum' in the Table I.
This is consistent with the earlier calculations on pion interferometry
~\cite{DHR} that the ratio $R_{out}/R_{side}$ will be larger 
than unity in a first
order phase transition scenario due to the appearance of mixed phase
and hence time delay due to the slow down of the expansion rate.
We would like to emphasize here that the HBT radii give the length
of homogeneity of the source~\cite{siny} and this is equal to the geometric 
size if the source is static. However, for a dynamic source 
{e.g.} the system formed after ultra-relativistic heavy ion 
collisions, the HBT radii is smaller than the geometric 
size (see ~\cite{chappm,csorgo,bertsch,xu}).

\bef
\begin{center}
\includegraphics[scale=0.4]{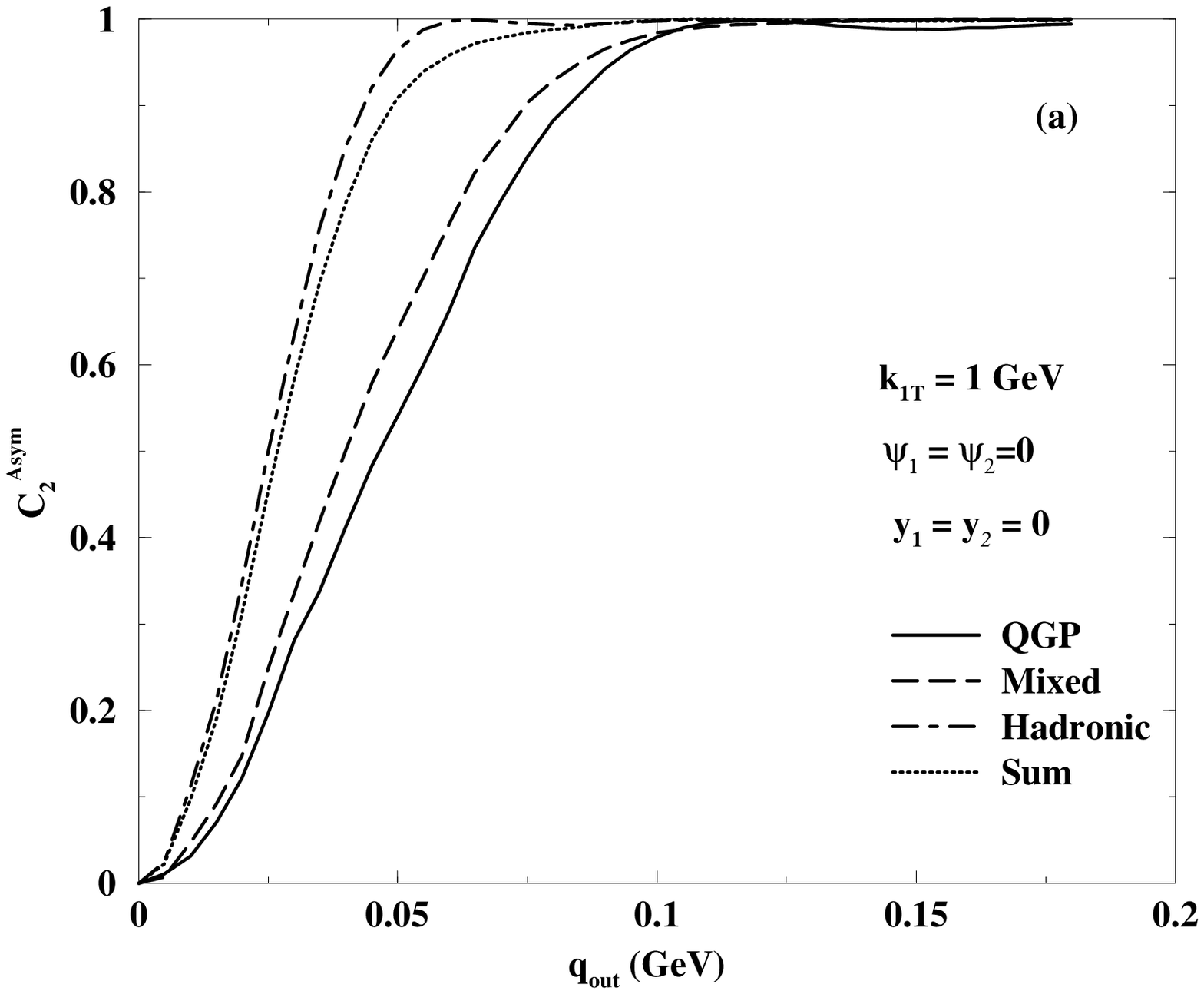}
\includegraphics[scale=0.4]{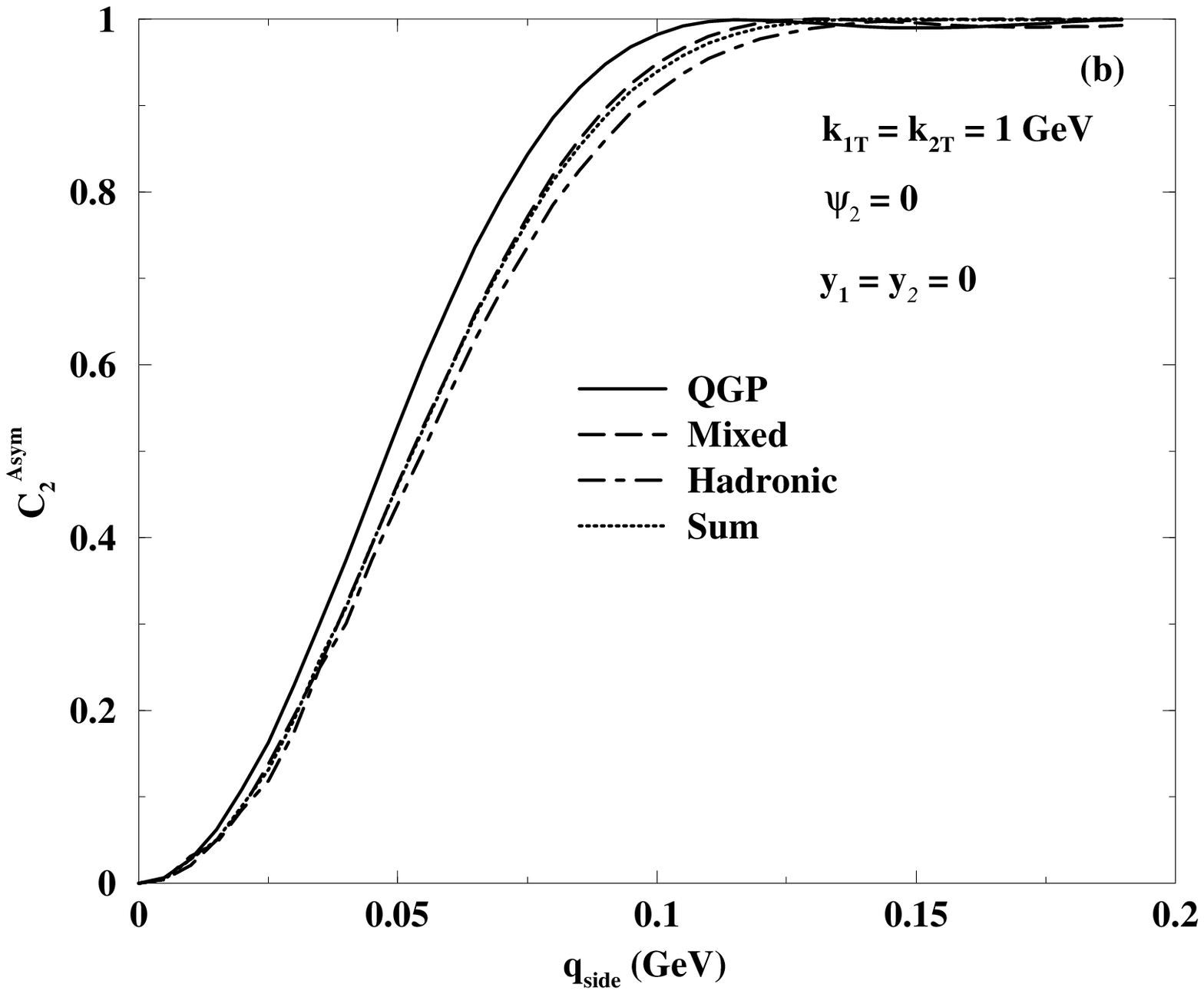}
\caption{ Same as Fig.~\ref{fig1}(b-c) when the mass
of the $\rho$ varies with temperature according to
BR scaling.
}
\label{fig2}
\end{center}
\eef
As indicated earlier, we also consider a scenario where the
$\rho$ meson mass varies with the temperature of the medium. The
obvious motivation is to comment on this very important issue based on  
two-electron interferometry. 
In Figs.~\ref{fig2}(a-b) we show the variation of the correlation 
strength ($C_{2}$) as a function of $q_{out}$ and $q_{side}$
for various phases with the 
masses of the $\rho$  modified in the medium according to BR
scaling. Unfortunately, these results are not very different from 
the corresponding ones without medium effects shown
in Fig.~\ref{fig1}. We do not present the results 
for $q_{long}$ and $q_{inv}$ as they are very similar to the case 
without mass variation. All these indicate that lepton interferometry
is not, probably, a suitable probe to detect the in-medium modification 
of vector mesons.

We also note that the HBT radii of various phases extracted from photon 
interferometry~\cite{photon} with spin-averaged 
source function and those obtained here from electron interferometry
with the spin-dependent source functions are similar.
This indicates that the results from
photon interferometry is not very sensitive to the 
spin of the photon. 

\renewcommand{\arraystretch}{1.5}
\begin{center}
\begin{tabular}{|c|c|c|c|c|c|c|}
\hline
& & $R_{inv}$ & $R_{out}$ & $R_{side}$ & $R_{long}$\\
&  &   &  & &     \\
\hline
& QGP & 3.5 & 3.5 & 3.5 & 0.6\\
RHIC & Mixed & 3.2 & 4.3 & 3.2 & 1.8 \\
& Hadron & 3.0 & 6.5 & 3.0 & 3.0 \\
& Sum & 3.2 & 6.0 & 3.2 & 3.3\\
\hline
\end{tabular}
\end{center}
Table 1 : Values of the various HBT radii in fm.

\section{SUMMARY and OUTLOOK}
The two-electron correlation functions have been evaluated for RHIC energies.  
The spin dependence of the source functions of 
the electrons originating from the QGP phase and hadronic
phase have been considered explicitly through the invariant amplitude.
(3+1) dimensional relativistic hydrodynamics has been used for the 
evolution in space and time from the initial QGP phase to the final
hadronic phase with intermediate mixed phase of QGP and hadrons 
in a first order phase transition scenario. 
In contrast to the
invariant mass distribution of the dileptons~\cite{annals,rapp},
the HBT radii obtained
from the two electrons interferometry is seen to be insensitive
to the in-medium modifications of the intermediary vector mesons,
apparently because of the cancellation of such effects between
the numerator and the denominator of Eq.~\ref{eq1}.
The results obtained from the electron interferometry here is
similar to those obtained from the photon interferometry~\cite{photon}
(where spin averaged amplitudes were used for photon production). 
The values of HBT radii extracted 
from the two-lepton correlation functions show that $R_{out}/R_{side}\,\sim 2$, 
consistent with the assumption of a first order phase transition 
scenario~\cite{DHR}. It has been checked that the HBT radii extracted
from the correlation function evaluated by using symmetric space
and anti-symmetric spin wave function is similar to the values 
given in Table 1. 
Experimentally it is difficult to verify the predictions
made here, however, first attempt to measure two photon correlation functions
in heavy ion collisions has already been made by WA98 collaboration~\cite{WA98}.

Unlike hadrons which are dominantly emitted from the freeze-out
surface of the fireball, the leptons and photons are produced and 
emitted from all the evolution stages of the matter formed after
the nuclear collisions. Therefore, lepton and photon interferometry
can give, in principle, `the length of homogeneity' of the system
at any stage of the evolution.
We have not come across any theoretical 
work on lepton interferometry with spin dependent 
invariant amplitude and  (3+1)dimensional hydrodynamical
expansion.  However, there are scopes of improvements of the 
present work.  
A first order phase transition is assumed here
in absence of satisfactory understanding of the order of the phase
transition from lattice QCD~\cite{karsch,fodor}. It will be interesting to
look into the two particle correlation functions with a continuous
transitions. Near the phase transition the quarks and gluons may
not behave as massless particles due its interactions with other
particles in the thermal bath although they are treated here as
massless. The bag model equation of state has been used here 
for simplicity, work with more realistic EOS~\cite{karsch}
is necessary. We have assumed that $e^-$ from annihilation
of thermal quarks and pions dominates at $p_T=1$ GeV. To 
justify this a detail analysis of the $p_T$ distribution
of $e^-$'s from other sources, {\it e.g.} Dalitz decays, open charm
decays, Drell-Yan processes etc are required.
A detailed calculations taking in to account
some of these factors will be published elsewhere.

{{\bf Acknowledgments}:  We would like to thank B. R. Schlei 
for useful discussions.
}

\end{document}